\documentclass[sort&compress,a4paper,oneside,3p,12pt]{melsarticle}
\usepackage{verbatim}
\usepackage{amssymb}
\usepackage{amsmath,amssymb,amsfonts,graphicx,float,multicol,fleqn}
\usepackage{multirow}
\numberwithin{equation}{section}
\newtheorem{theorem}{Theorem}[section]

\begin{document}
\begin{frontmatter}
\title{Comparison between two common collocation approaches based on radial basis
functions for the case of heat transfer equations arising in porous medium}


\author[a,b]{K. Parand}
\ead{k\_parand@sbu.ac.ir}
\fntext[a]{Member of research group of Scientific Computing.}
\author[c]{S. Abbasbandy\corref{cor}}
\ead{abbasbandy@yahoo.com} 
\cortext[cor]{Corresponding author. Tel:(+98912) 1305326 Fax:(+98281) 3780040}
\author[c]{S. Kazem}
\ead{saeedkazem@gmail.com}
\author[b]{A.R. Rezaei}
\ead{alireza.rz@gmail.com}
\address[b]{Department of Computer Sciences, Shahid Beheshti University, G.C., Tehran, Iran}
\address[c]{Department of Mathematics, Imam Khomeini International University, Ghazvin 34149-16818, Iran}

\begin{abstract}
In this paper two common collocation approaches based on radial
basis functions have been considered; one be computed through the
integration process (IRBF) and one be computed through the
differentiation process (DRBF). We investigated the two approaches
on natural convection heat transfer equations embedded in porous
medium which are of great importance in the design of canisters
for nuclear wastes disposal. Numerical results show that the IRBF
be performed much better than the common DRBF, and show good
accuracy and high rate of convergence of IRBF process.\\
\end{abstract}

\begin{keyword}
Collocation method; Nonlinear ODE; Radial Basis Functions; Direct
Inverse Multiquadric; Indirect Multiquadric; Porous media.
\PACS 47.56.+r, 02.70.Hm
\end{keyword}
\end{frontmatter}
\section{Introduction}\label{Introduction}
Natural convective heat transfer in porous media has received
considerable attention during the past few decades. This interest
can be attributed due to its wide range of applications in ceramic
processing, nuclear reactor cooling system, crude oil drilling,
chemical reactor design, ground water pollution and filtration
processes. External natural convection in a porous medium adjacent
to heated bodies was analyzed by Nield and Bejan \cite{ Bejan and
Nield book}, Merkin \cite{Merkin 1978, Merkin 1979}, Minkowycz and
Cheng \cite{Minkowycz and Cheng 1982, Minkowycz and Cheng 1985},
Pop and Cheng \cite{Cheng and Pop, Pop and Cheng}, Ingham and Pop
\cite{Ingham and Pop}. In all of these analysis, it was assumed
that boundary layer approximations are applicable and the coupled
set of governing equations were solved by numerical methods.

\par
In this paper, the same approximations are applied to the problem
of natural convection about an inverted heated cone embedded in a
porous medium of infinite extent. No similarity solution exists
for the truncated cone, but for the case of full cone, if the
prescribed wall temperature or surface heat flux is a power
function of distance from the vertex of the inverted cone
similarity solutions exist \cite{Bejan and Nield book,Cheng and
Pop}, a great deal of information is available on heat and fluid
flow about such cones as reviewed by Refs. \cite{Pop I Ingham DB
2001,Vafai K 2000}.

\par
Bejan and Khair \cite{Bejan and khair} used Darcy's law to study
the vertical natural convective flows driven by temperature and
concentration gradients. Nakayama and Hossain \cite{Nakayama and
Hossain} applied the integral method to obtain the heat and mass
transfer by free convection from a vertical surface with constant
wall temperature and concentration. Yih \cite{Yih 1999
truncated_cone} examined the coupled heat and mass transfer by
free convection over a truncated cone in porous media for variable
wall temperature and concentration or variable heat and mass
fluxes and \cite{Yih 1999} applied the uniform transpiration
effect on coupled heat and mass transfer in mixed convection about
inclined surfaces in porous media for the entire regime. Cheng
\cite{Cheng 2000} used an integral approach to study the heat and
mass transfer by natural convection from truncated cones in porous
media with variable wall temperature and \cite{Cheng 2009} studied
the Soret and Dufour effects on the boundary layer flow due to
natural convection heat and mass transfer over a vertical cone in
a porous medium saturated with Newtonian fluids with constant wall
temperature. Natural convective mass transfer from upward-pointing
vertical cones, embedded in saturated porous media, has been
studied using the limiting diffusion \cite{Rahman 2007}. The
natural convection along an isothermal wavy cone embedded in a
fluid-saturated porous medium are presented in \cite{Pop and Na
1994, Pop and Na 1995}. Lai and Kulacki \cite{Lai and Kulacki}
studied the natural convection boundary layer flow along a
vertical surface with constant heat and mass flux including the
effect of wall injection. In \cite{Sohouli.Famouri} fluid flow and
heat transfer of vertical full cone embedded in porous media have
been solved by Homotopy analysis method
\cite{Abbasbandy.Hayat.CNSNS2009,Abbasbandy.CEJ2008}.

Mathematical modeling of many problems in science and engineering
leads to ordinary differential equations (ODEs) \cite{Parand.Phys.
Scripta2004,
Parand.Dehghan.Sinc.Blas,Parand.Dehghan.Rezaei.CPC,Parand.Rezaei.Ghaderi.CNSNS,Abbasbandy.ShivanianPLA2010}.
The methods based on radial basis functions (RBF) which are part
of an emerging field of mathematics are famous ways to solve these
kinds of problems . First studied by Roland Hardy, an Iowa
State geodesist, in 1968, these methods allow for scattered data
to be easily used in computations \cite{N. Mai-Duy 2005}. The
concept of solving DEs by using RBFs was first introduced by Kansa
\cite{Kansa EJ 1990}. Since then, it has received a great deal of
attention from researchers. And consequently, many further interesting
developments and applications have been reported (e.g. Zerroukat
et al.\cite{Zerroukat M 1998}, Mai-Duy and Tran-Cong\cite{Mai-Duy
N 2001,Tran-Cong 2001}). Essentially, in a typical RBF collocation
method, each variable and its derivatives are all expressed as
weighted linear combinations of basis functions, where the sets of
network weights are identical. These closed forms of representations
are substituted with the governing equations as well as boundary
conditions, and the point collocation technique is then employed
to discretize the system \cite{Parand.Rezaei.Ghaderi.CNSNS}. If
all basis functions in networks are available in analytic forms,
the RBF collocation methods can be regarded as truly meshless
methods \cite{Bengt Fornberg 2008}. There are two basic approaches
for obtaining new basis functions from RBFs, namely direct
approach (DRBF) based on a differential process (Kansa \cite{Kansa
EJ 1990}) and indirect approach (IRBF) based on an integration
process (Mai-Duy and Tran-Cong \cite{N. Mai-Duy 2005,Mai-Duy N
2001,Mai-Duy N.Tran-Cong T 2003}). Both approaches were tested on
the solution of second order DEs and the indirect approach was
found to be superior to the direct approach (Mai-Duy and Tran-Cong
\cite{Mai-Duy N 2001}).

In this paper we apply the DRBF and IRBF for solving natural convection of Darcian
fluid about a vertical full cone embedded in porous media prescribed
surface heat flux which is third order nonlinear ODE.


\section{Problem formulation}
\begin{figure}
\includegraphics[scale=0.7]{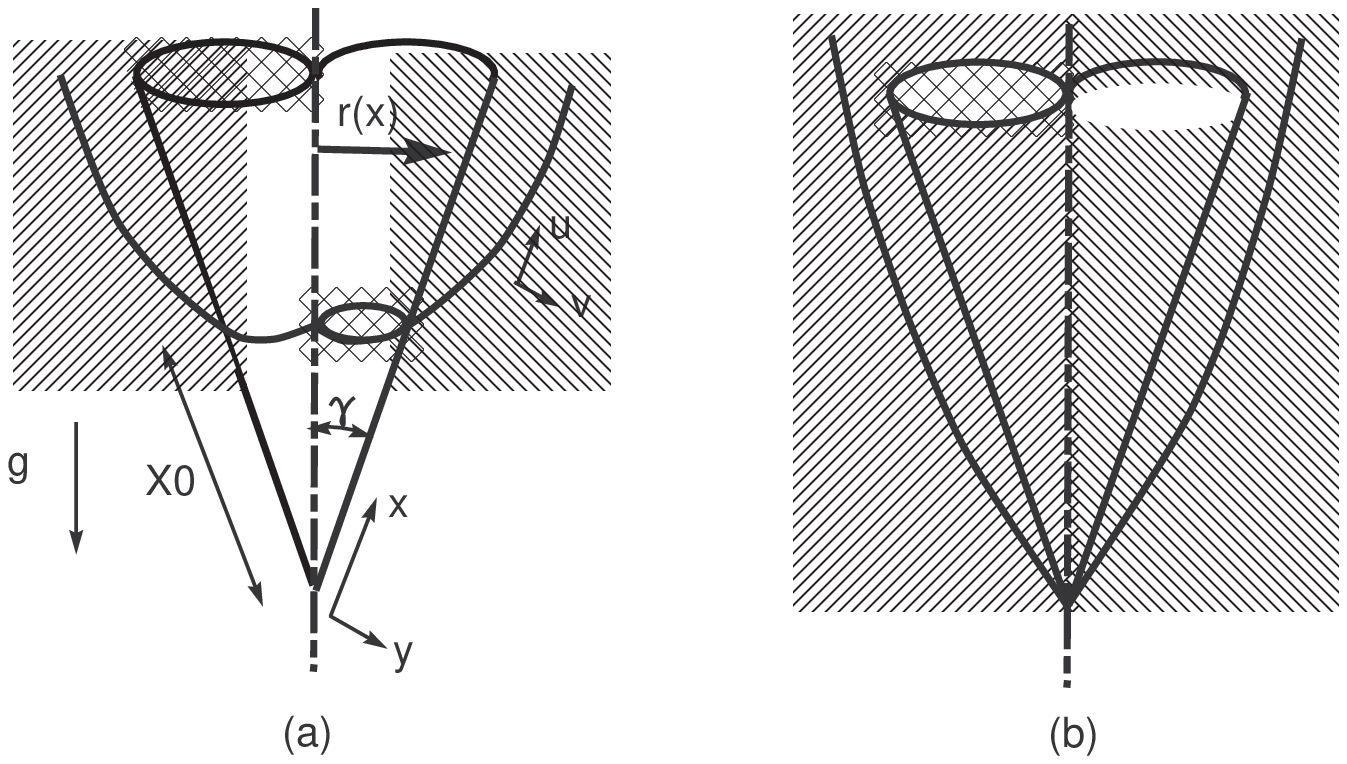}
\caption{(a) Coordinate system for the boundary layer on a heated
frustum of a cone, (b) full cone, $x_0=0$.}
\label{Fig.FullCone}
\end{figure}

Consider an inverted cone with semi-angle $\gamma$ and take axes
in the manner indicated in Fig. \ref{Fig.FullCone}(a). The
boundary layer develops over the heated frustum $x=x_0$.

The boundary layer equations for natural convection of Darcian
fluid about a cone are \cite{Cheng and Pop}:
\begin{eqnarray}
&&\frac{\partial}{\partial x}(ru)+\frac{\partial}{\partial y}(rv)=0,\\
&&u=\frac{\rho_\infty \beta K g \cos\gamma (T-T_\infty)}{\mu},\\
&&u\frac{\partial T}{\partial x}+v\frac{\partial T}{\partial y}=
\alpha \frac{\partial^2 T}{\partial y^2}.\nonumber
\end{eqnarray}
For a thin boundary layer, $r$ is obtained approximately $x sin(\gamma)$.
Suppose that a power law of heat flux is prescribed on the frustum. Accordingly,
the boundary conditions at infinity are:
\begin{eqnarray}
u=0,&~T = T_\infty,~~\text{if}~~y\rightarrow\infty\\
\end{eqnarray}
and at the wall are
\begin{eqnarray}
v=0~~\text{if}~~ y=0. \nonumber
\end{eqnarray}

If the surface heat flux $q_w$ \cite{Cheng and Pop} is prescribed, $q_w$ is obtained as
\begin{eqnarray}
q_w= {- k (\frac{\partial T}{\partial
y}})_{y=0}=A(x-x_0)^{\lambda},~~~~~x_0 \leq x \leq \infty.
\nonumber
\end{eqnarray}
For the case of a full cone $(x_0=0, Fig. \ref{Fig.FullCone}(b))$ a similarity solution exists \cite{Cheng and Pop}.
\par

In the case of prescribed surface heat flux the similarity solution for the stream function $\psi$ and $T$ where
\begin{eqnarray}
u = \frac{1}{r}\frac{\partial \psi}{\partial y},~~~
v = -\frac{1}{r}\frac{\partial \psi}{\partial x}
\end{eqnarray}
is of the form \cite{Cheng and Pop}:
\begin{eqnarray}\label{wall temperature1 1}
&&\psi = \alpha r (R a_x)^{1/3}f(\eta),~~~~~~~~\\
&&T - T_\infty = \frac{q_w x}{k}(Ra_x)^{-\frac{1}{3}}\theta(\eta), \nonumber \\
&&\eta = \frac{y}{x}(Ra_x)^{1/3}, \nonumber
\end{eqnarray}
where
\begin{eqnarray}\label{wall temperature2 1}
Ra_x = \frac{\rho_\infty \beta g K \cos(\gamma) q_w x^2}{\mu \alpha k}
\end{eqnarray}
is the local Rayleigh number for the case of prescribed surface heat flux.
The governing equations become
\begin{eqnarray}\label{energy equations1}
f' = \theta, ~~~~~~~~~~~~~~~~~~~~~~~~~~~~~~~~\\
\theta'' + \frac{\lambda + 5}{2}f\theta'- \frac{2\lambda + 1}{3} f' \theta = 0,\nonumber
\end{eqnarray}
subjected to boundary conditions as:
\begin{equation}\label{boundary conditions}
f(0) = 0, ~~~\theta' (0) = -1, ~~~\theta (\infty) = 0.
\end{equation}
Finally from Eqs.~(\ref{energy equations1}) and (\ref{boundary
conditions}) we have:
\begin{align}\label{MainEquation}
\begin{cases}
ODE. \quad f'''+ \left(\frac{\lambda +
5}{2}\right)ff''-\left(\frac{2\lambda+1}{3}\right)(f')^2 = 0, \cr
B.C. \quad f(0)=0, \quad f''(0)=-1, \quad f'(\infty) = 0.
\end{cases}
\end{align}
It is of interest to obtain the value of the local Nusselt number which is defined as \cite{Cheng and Pop}:
\begin{eqnarray}\label{Nusselt number}
Nu_x=\frac{q_w x}{k(T_w-T_\infty)}.
\end{eqnarray}
From Eqs.~(\ref{Nusselt number}), (\ref{wall temperature1 1}) and
(\ref{wall temperature2 1}) it follows that the local Nusselt
number which is interest to obtain given by:
\begin{eqnarray}
Nu_x=Ra_x^{1/3}[-\theta (0)].
\end{eqnarray}

\section{RBF Functions}\label{RCF.Intro}
Let $\mathbb{R}^{+}=\{x\in \mathbb{R},x\geq 0\}$ be the non-negative half-line and let $\phi:\mathbb{R}^{+}\to \mathbb{R}$ be a continuous function with $\phi(0)\geq 0$. A radial basis function on $\mathbb{R}^{d}$ is a function of the form
\begin{equation}\nonumber
    \phi(\|X-X_i\|)
\end{equation}
where $X, ~X_i \in \mathbb{R}^{d}$ and $\|.\|$ denotes the Euclidean distance between $X, ~X_i$. If one chooses $N$ points $\{X_i\}_{i=1}^{N}$ in $\mathbb{R^{d}}$ then by custom
\begin{equation}\nonumber
s(X)=\sum_{i=1}^{N}\lambda_i\phi(\|X-X_i\|);\quad \lambda_i \in
\mathbb{R},
\end{equation}
is called a radial basis function as well \cite{golberg}.\\
In order to explain RBF methods briefly, suppose that the one-dimensional input data point set or the center
set $x_i$ in the given domain $\Omega \subseteq \mathbb{R}$ is given. The center point is not necessarily structured, that is, it can have an arbitrary distribution. The arbitrary grid structure
is one of the major differences between the RBF method and other global methods. Such a mesh-free grid structure
yields high flexibility especially when the domain is irregular. In this work the uniform grid is used for RBF approximation.

\subsection{Properties of RBF}
With a radial function $\phi(r )$ and with data values $u_{i}$ given at the locations $x_{i}$, for $i=1,2,...,N$ the function \\
\begin{equation}\label{rbf.intro}
s(x)=\sum_{i=1}^{N}\lambda_{i}\phi_{i}(x)
\end{equation}
where $r=r_{i}=\|x-x_{i}\|$ and $\phi_{i}(x)=\phi(\|x-x_{i}\|)$, interpolates the
data if we choose the expansion coefficients $\lambda_{i}$ in such
a way that $s(x_{j})=u_{j}$, for $j=1,2,...,N$ \cite{M.D.Buhmann
2000,Fasshauer.G.E.(2007)}. The expansion
coefficients $\lambda_{i}$ can
therefore be obtained by solving the linear system $A\Lambda=U$, where:
\begin{eqnarray}\label{rbf.mat}
A_{ij}=\phi(\|x_{j}-x_{i}\|),\\
\Lambda=[\lambda_{1},\lambda_{2},...,\lambda_{N}]^T,\\
U=[u_1, u_2, \dots, u_N]^T.
\end{eqnarray}

All the infinitely smooth RBF choices listed in
Table~(\ref{Tab.RBR.definition}) will give coefficient matrices
$A$ in (\ref{rbf.mat}) which are symmetric and nonsingular
\cite{Powell}, i.e. there is a unique interpolant of the form
(\ref{rbf.intro}) no matter how the distinct data points are
scattered in any number of space dimensions. In the cases of
inverse quadratic, inverse multiquadric and GA the matrix $A$ is
positive definite and, for multiquadric (MQ), it has one positive
eigenvalue and the
remaining ones are all negative \cite{Powell}.\\
Interpolation using Conical splines and thin-plate splines (TPSs) can
become singular in multidimensions \cite{Bengt Fornberg 2008}.
However, low-degree polynomials can be added to the RBF
interpolant to guarantee that the interpolation matrix is positive
definite (a stronger condition than nonsingularity). For example,
for the Conical RBF and the TPS in $d$ dimensions this becomes the case if
we use as an interpolant
$s(x)=\sum_{i=1}^{m}a_{i}p_{i}(x)+\sum_{i=1}^{N}\lambda_{i}\phi(\|x-x_{i}\|)$
together with the constraints
$\sum_{j=1}^{N}\lambda_{j}p_{i}(x_j)=0$, for $i=1,2,...,m$. Here
$p_{i}(x)$ denotes a basis for polynomials of $\mathcal{P}^d_{q}$
in $\mathbf{R}^d$ ($\mathcal{P}^d_{q}$ denotes the space of
d-variate polynomials of order not exceeding q) and $m=(q-1+d)!/(d!(q-1)!)$ \cite{Mehdi Dehghan
· Ali Shokri}.
\subsection{RBF Interpolation}
One dimensional function $u(x)$ to be interpolated or approximated can be
represented by an RBF as:
\begin{eqnarray}\label{rbf intrerpolation}
u(x)\approx s(x)=\sum_{i=1}^{N}\lambda_{i}\phi_{i}(x)=\Phi^T(x)\Lambda
\end{eqnarray}
where
\begin{eqnarray}
&&\Phi^T(x)=[\phi_{1}(x),\phi_{2}(x),...,\phi_{N}(x)],\\
&&\Lambda=[\lambda_{1},\lambda_{2},...,\lambda_{N}]^T,
\end{eqnarray}
$x$ is the input and $\lambda_{i}$s are the set of coefficients to be determined. By choosing $N$ interpolate nodes $(x_{j},j=1,2,...,N)$ in  $\Omega\cup\partial\Omega$, the function $u(x)$ can be approximated in $\Omega\cup\partial\Omega$.
\begin{equation}
u_{j}=\sum_{i=1}^{N}\lambda_{i}\phi_{i}(x_{j})  , \quad
(j=1,2,...,N).
\end{equation}
To brief discussion on coefficient matrix we define:
\begin{equation}
A\Lambda=U
\end{equation}
where
\begin{align}
U=&[u_{1},u_{2},...,u_{N}]^T,\\\nonumber
A=&[\Phi^T(x_{1}),\Phi^T(x_{2}),...,\Phi^T(x_{N})]^T\\
 =&
\left(
 \begin{array}{cccc}\label{matrice}
\phi_{1}(x_{1}) & \phi_{2}(x_{1}) & \dots  & \phi_{N}(x_{1}) \\
\phi_{1}(x_{2}) & \phi_{2}(x_{2}) & \dots  & \phi_{N}(x_{2}) \\
\vdots          & \vdots          & \ddots & \vdots          \\
\phi_{1}(x_{N}) & \phi_{2}(x_{N}) & \dots  & \phi_{N}(x_{N}) \\
\end{array}
\right).
\end{align}

Note that $\phi_{i}(x_{j})=\phi(\|x_{i}-x_{j}\|)$ therefore $\phi_{i}(x_{j})=\phi_{j}(x_{i})$ consequently $A=A^T$.\\
The shape parameter $c$ which is shown in
Table~(\ref{Tab.RBR.definition}) affects both the accuracy of the
approximation and the conditioning of the interpolation matrix
\cite{S Sarra}. In general, for a fixed number of $N$, smaller
shape parameters produce the more accurate approximations, but
also are associated with a poorly conditioned $A$. The condition
number also grows with $N$ for fixed values of the shape parameter
$c$. Many researchers (e.g.\cite{Carlson and Foley,Rippa}) have
attempted to develop algorithms for selecting optimal values of
the shape parameter. The optimal choice of the shape parameter is
still an open question. In practice it is most often selected by
brute force. Recently, Fornberg et. al.\cite{B. Fornberg} have
developed a Contour-Pad\'{e} algorithm which is
capable of stably computing the RBF approximation for all $c>0$ \cite{S Sarra}.\\
The following theorem about the convergence of RBF
interpolation is discussed \cite{WuZM 2002,WuZM 1993}.
\begin{theorem}
assume ${x_{i}},(i=1,2,...,N)$ are N nodes in $\Omega$ which is convex, let\\$$h=\max_{x\in\Omega}\min_{1\leq i\leq N} \|x-x_{i}\|_{2}$$\\
when $\hat{\phi}(\eta)<c(1+|\eta|)^{-(2l+d)}$ for any u(x)
satisfies$ \int{(\hat{u}(\eta))^2}/{\hat{\phi}(\eta)}
d\eta<\infty$ we have
$$\|u^{N(\alpha)}-u^{(\alpha)}\|_{\infty}\leq{ch^{l-\alpha}}$$ where
$\phi(x)$ is RBF and the constant $c$ depends on the RBF, d is
space dimension, l and $\alpha$ are nonnegative integer. It can
be seen that not only RBF itself but also its any order
derivative has a good convergence.
\end{theorem}

\subsection{Direct RBF for ODEs (DRBF)}
In the direct method, the closed form DRBF approximating function
(\ref{rbf intrerpolation}) is first obtained from a set of training
points, and its derivative of any order, e.g. $p$th order, can then
be calculated in a straightforward manner by differentiating such
a closed form DRBF as follows:
\begin{eqnarray}\label{RBFC}
   \frac{d^ks(x)}{dx^k}=\frac{d^k}{dx^k}(\sum_{i=1}^{N}\lambda_{i}\phi_{i}(x))=\sum_{i=1}^{N}\lambda_{i}
   \frac{d^k\phi_{i}(x)}{dx^k}=\sum_{i=1}^{N}\lambda_{i}G_{i}^{[k]}(x)
\end{eqnarray}
where \begin{eqnarray}\nonumber
 G_{i}^{[k]}(x)={d^k\phi_{i}(x)}/{dx^k}, \qquad k=0,1,...,p.
 \end{eqnarray}
Now we aim to apply the DRBF method for solving the ODEs in general form :
\begin{eqnarray}\label{initial ode}
\begin{cases}
F(x,u,u',...,u^{(p-1)},u^{(p)})=0,& a \leq x\leq b, \cr
u^{(i)}(e_{i})=\alpha_{i+1},& i=0,1,...,p-1,
\end{cases}
\end{eqnarray}
where $e_{i}\in\{a,b\}$ and  $u^{(i)}(x)=d^iu(x)/dx^i$, $F$ is
known function and $\{\alpha_{i}\}_{i=1}^{p}$ are known constants.
By substituting Eq.~(\ref{rbf intrerpolation}) in (\ref{initial
ode}) and using Eq.~(\ref{RBFC}) we have:
\begin{eqnarray}\nonumber
   F(x,s,s',...,s^{(p)})=F(x,\sum_{i=1}^{N}\lambda_{i}G_{i}^{[0]}(x),\sum_{i=1}^{N}\lambda{i}G_{i}^{[1]}(x),...,
   \sum_{i=1}^{N}\lambda_{i}G_{i}^{[p]}(x)).
\end{eqnarray}
Now, to obtain $\lambda_i$s $(i=1,2,...,N)$
we define the residual function:
\begin{eqnarray}\label{Res}
    Res(x)=F(x,s,s',...,s^{(p)}).
\end{eqnarray}

The set of equations for obtaining the coefficients
$\{\lambda_{i}\}_{i=1}^{N}$ come from equalizing Eq.~(\ref{Res})
to zero at $N-p$ interpolate nodes $\{x_{j}\}_{j=1}^{N-p}$ plus
$p$ boundary conditions:
\begin{equation}
\begin{cases}
Res(x_{j})=0, & j=1,2,...,N-p,\cr
\sum_{i=1}^{N}\lambda_{i}G_{i}^{[k]}(e_{i})=\alpha_{i+1},
&k=0,1,...,p-1.
\end{cases}
\end{equation}
 Since the direct approach is based on a differentiation process,
all derivatives obtained here are very sensitive to noise arising
from the interpolation of DRBFs from a set of discrete data
points. Any noise here, even at the small level, will be badly
magnified with an increase in the order of derivative \cite{N. Mai-Duy 2005}.
\subsection{Indirect RBF for ODEs (IRBF)}
In the indirect method, the formulation of the problem starts with
the decomposition of the highest order derivative under
consideration into RBFs. The obtained derivative expression is
then integrated to yield expressions for lower order derivatives
and finally for the original function itself. In contrast, the
integration process, where each integral represents the area under
the corresponding curve, is much less sensitive to noise. Based on
this observation, it is expected that through the integration
process, the approximating functions are much smoother and
therefore have higher approximation power. Also To numerically explore tile IRBF methods with shape parameters for which the interpolation matrix is too poorly conditioned to use standard methods \cite{Fornberg and Wright}. Let $p$ be the highest
order of the derivative under consideration the boundary value
ODEs in general form Eq.~(\ref{initial ode}) when ($\exists k$
s.t. $e_{k}=b$) then we can define:
\begin{align}\label{irbf.b}
\frac{d^p\hat{s}(x)}{dx^p}&=\sum_{i=1}^{N}\lambda_{i}\phi_{i}(x),\\\nonumber
\frac{d^{p-1}\hat{s}(x)}{dx^{p-1}}&=\int{\sum_{i=1}^{N}\lambda_{i}\phi_{i}(x)dx}=\sum_{i=1}^{N}
\lambda_{i}\int{\phi_{i}(x)}dx=\sum_{i=1}^{N}\lambda_{i}h_{i}^{[p-1]}(x)+d_{1},\\\nonumber
\vdots\\\nonumber
\frac{d\hat{s}(x)}{dx}&=\int{\sum_{i=1}^{N}\lambda_{i}h_{i}^{[2]}(x)}dx+d_{1}\frac{x^{p-2}}{(p-2)!}+d_{2}
\frac{x^{p-3}}{(p-3)!}+...+d_{p-1}\\\nonumber
&=\sum_{i=1}^{N}\lambda_{i}h_{i}^{[1]}(x)+d_{1}\frac{x^{p-2}}{(p-2)!}+d_{2}
\frac{x^{p-3}}{(p-3)!}+...+d_{p-1},\\\nonumber
\hat{s}(x)&=\int({\sum_{i=1}^{N}\lambda_{i}h_{i}^{[1]}(x)}+d_{1}
\frac{x^{p-2}}{(p-2)!}+d_{2}\frac{x^{p-3}}{(p-3)!}+...+d_{p-1})dx\\\nonumber
&=\sum_{i=1}^{N}\lambda_{i}h_{i}^{[0]}(x)+d_{1}\frac{x^{p-1}}{(p-1)!}+d_{2}\frac{x^{p-2}}{(p-2)!}+...+d_{p-1}x+d_{p},
\end{align}
where\\
\begin{equation}
h_{i}^{[k]}(x)=
\begin{cases}
\int\phi_{i}(x)dx ,&k=p-1, \cr \int
h_{i}^{[k+1]}(x)dx,&k=0,1,...,p-2.
\end{cases}
\end{equation}
Substituting Eqs.~(\ref{irbf.b}) in (\ref{Res}) at $N$ interpolate
nodes $\{x_{j}\}_{j=1}^{N}$
 plus $p$ boundary conditions the set of coefficients $\{\lambda_{i}\}_{i=1}^{N}$ and $\{d_{j}\}_{j=1}^{p}$ is obtained as follow :\\
\begin{eqnarray}\nonumber
  \begin{cases}
    Res(x_{j})=0 ,\quad &j=1,2,...,N, \cr
    \hat{s}^{(i)}(e_{i})=\alpha_{i+1},\quad &i=1,2,...,p.
  \end{cases}
\end{eqnarray}

\section{Solving the model}\label{dd}
Consider governing equation of fluid flow and heat transfer of
full cone embedded in porous medium that is expressed by
Eq.~(\ref{MainEquation}) for prescribed surface heat flux.

In the first step of our analysis, we approximate $f(\eta)$ for solving the model by DRBF:
\begin{align}\label{LDRBF}
f(\eta)\simeq
s(\eta)&=\sum_{i=1}^{N}\lambda_{i}\phi_{i}(\eta),\\\nonumber
\end{align}
and $f'''(\eta)$ for solving the model by IRBF:
\begin{align}\label{IIRBF}
f'''(\eta)\simeq\frac{d^{3}}{d\eta^3}\hat{s}(\eta)&=\sum_{i=1}^{N}\lambda_{i}\phi_{i}(\eta).
\end{align}
The general form of problem appear to:
\begin{align}
 \begin{cases}\label{EQQQ}
      F_{\lambda}(\eta,f,f',f'',f''')=0,\cr
      f(0)=0,\ f''(0)=-1,\ f'(\infty)=0.
 \end{cases}
\end{align}
To solve this problem we define residual function:
\begin{align}
 \label{RRR}
      Res(\eta)=F_{\lambda}(\eta,s,s',s'',s''');\quad for\ DRBF,
\end{align}
\begin{align}\label{RRRI}
Res(\eta)=F_{\lambda}(\eta,\hat{s},\hat{s}',\hat{s}'',\hat{s}''');
\quad for\ IRBF.
\end{align}
The unknown coefficients $\{\lambda_i\}_{i=1}^{N}$ come from
equalizing $Res(\eta)$ to zero at $N$ interpolate nodes $\eta_{i}$
from Uniform distribution between $0$ and $\eta_\infty$ which we
set $9/2$ for this problem.

\subsection{Solving the model by DRBF}
\begin{figure}
\includegraphics[scale=0.7]{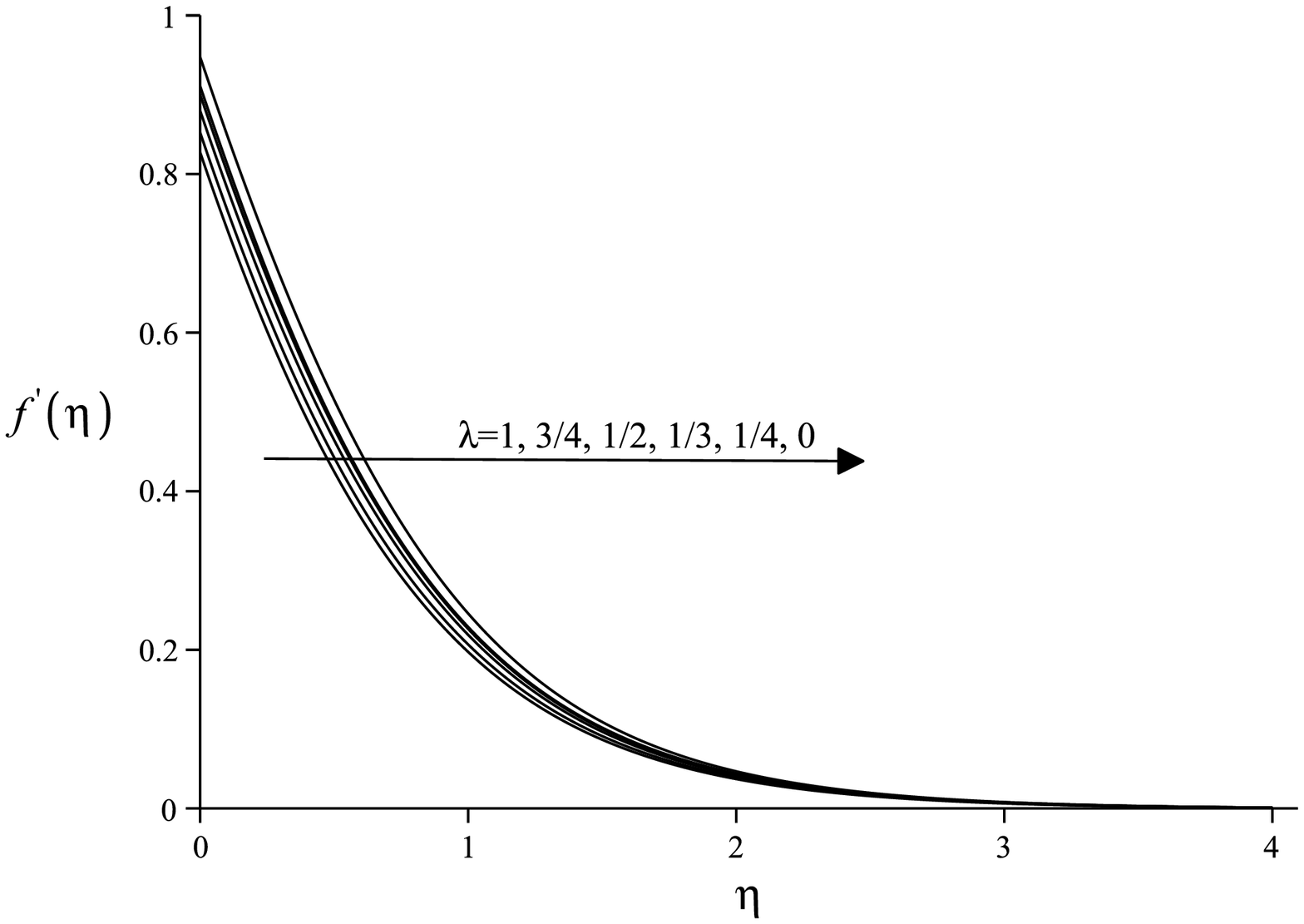}
\caption{DRBF approximation of $f'(\eta)$ for different values $\lambda=0,~1/4,~1/3,~1/2,~3/4$ and $1$}
\label{Fig.heat_flux_DRBF}
\end{figure}

In the first step of solving, $\phi_{i}(\eta)$ is set by inverse
multiquadric function which is shown in
Table~(\ref{Tab.RBR.definition}). Now, the residual
function is constructed by substituting Eq.~(\ref{LDRBF}) in Eq.~(\ref{RRR}):
\begin{align}
Res(\eta)&=\sum_{i=1}^{N}\lambda_{i}\phi_{i}^{'''}(\eta)+(\frac{\lambda+5}{2})
(\sum_{i=1}^{N}\lambda_{i}\phi_{i}(\eta))(\sum_{i=1}^{N}\lambda_{i}\phi_{i}^{''}(\eta))-
(\frac{2\lambda+1}{3})(\sum_{i=1}^{N}\lambda_{i}\phi_{i}^{'}(\eta))^2.\nonumber
\end{align}
By using $N-2$ interpolate nodes $\{\eta_{j}\}_{j=1}^{N-2}$ plus
two boundary conditions of Eq. (\ref{EQQQ}) $(f(0)=0,f''(0)=-1)$,
the set of equations can be solved, consequently  the coefficients
$\{\lambda_{i}\}_{i=1}^{N}$ will be obtained:
\begin{align}
  \begin{cases}
  Res(\eta_{j})=0, &\quad j=1,2,...,N-2,\cr
  \sum_{i=1}^{N}\lambda_{i}\phi_{i}(0)=0,\nonumber\cr
  \sum_{i=1}^{N}\lambda_{i}\phi_{i}''(0)=-1.\nonumber\cr
  \end{cases}
\end{align}
Take into account $\phi'_i(\infty)=0$, for $i=1,2,...,N$ the infinity boundary condition ($f'(\infty)=0$) is already satisfied.\\
Table~(\ref{Tab.DRBF.Resultsf'(0)}) show the $f'(\eta)$ for some
$\lambda$ in comparison with solutions of \cite{Sohouli.Famouri}.
Also $f'(\eta)$ for two selected $\lambda=1/4$ and $3/4$ are
showed in Table~(\ref{Tab.DRBF.Results.f'(etta)}) in comparison with Runge-Kutta solution is obtained by the MATLAB software command ODE45 which is used and applied by the authors in ref. \cite{Sohouli.Famouri}. Absolute errors
show that DRBF give us approximate solution with a high degree of
accuracy with a small $N$. The resulting graph of
Eq.~(\ref{MainEquation}) is shown in Figure
(\ref{Fig.heat_flux_DRBF}).

\subsection{Solving the model by IRBF}
In the first of solving $\phi_{i}(\eta)$ is set by multiquadric
function which is shown in Table~(\ref{Tab.RBR.definition}).
Now, the residual function is constructed by substituting
Eq.~(\ref{IIRBF}) in Eq.~(\ref{RRRI}) and using
Eq.~(\ref{irbf.b}):
\begin{align}\nonumber
Res(\eta)&=\sum_{i=1}^{N}\lambda_{i}\phi_{i}(\eta)+(\frac{\lambda+5}{2})\sum_{i=1}^{N}\lambda_{i}\int\int\int
\phi_{i}(\eta)d\eta\sum_{i=1}^{N}\lambda_{i}\int\phi_{i}(\eta)d\eta\\\nonumber
&-(\frac{2\lambda+1}{3})(\sum_{i=1}^{N}\lambda_{i}\int\int\phi_{i}({\eta})d\eta)^2\\\nonumber
&=\sum_{i=1}^{N}\lambda_{i}\phi_{i}(\eta)+(\frac{\lambda+5}{2})(\sum_{i=1}^{N}\lambda_{i}h_{i}^{[0]}(\eta)+
\frac{d_{1}\eta^2}{2}+d_{2}\eta+d_{3})(\sum_{i=1}^{N}\lambda_{i}h_{i}^{[2]}(\eta)+d_{1})\\\nonumber
&-(\frac{2\lambda+1}{3})(\sum_{i=1}^{N}\lambda_{i}h_{i}^{[1]}(\eta)+d_{1}\eta+d_{2})^2.
\end{align}
By using $N$ interpolate nodes $\{\eta_{j}\}_{j=1}^{N}$ plus three boundary conditions of Eq. (\ref{EQQQ})
the set of equations can be solved, consequently  the coefficients $\{\lambda_{i}\}_{i=1}^{N}$ and
$\{d_{i}\}_{i=1}^{3}$ will be obtained. In this method we put $\eta_\infty$ instead of infinity condition:
\begin{align}\nonumber
  \begin{cases}
     Res(\eta_{j})=0,&\quad j=1,2,...,N,\cr
     \sum_{i=1}^{N}\lambda_{i}h_{i}^{[0]}(0)+d_{3}=0,\cr
     \sum_{i=1}^{N}\lambda_{i}h_{i}^{[2]}(0)+d_{1}=-1,\cr
     \sum_{i=1}^{N}\lambda_{i}h_{i}^{[1]}(\eta_{\infty})d_{1}\eta_{\infty}+d_{2}=0.
  \end{cases}
\end{align}

\begin{figure}
\includegraphics[scale=0.7]{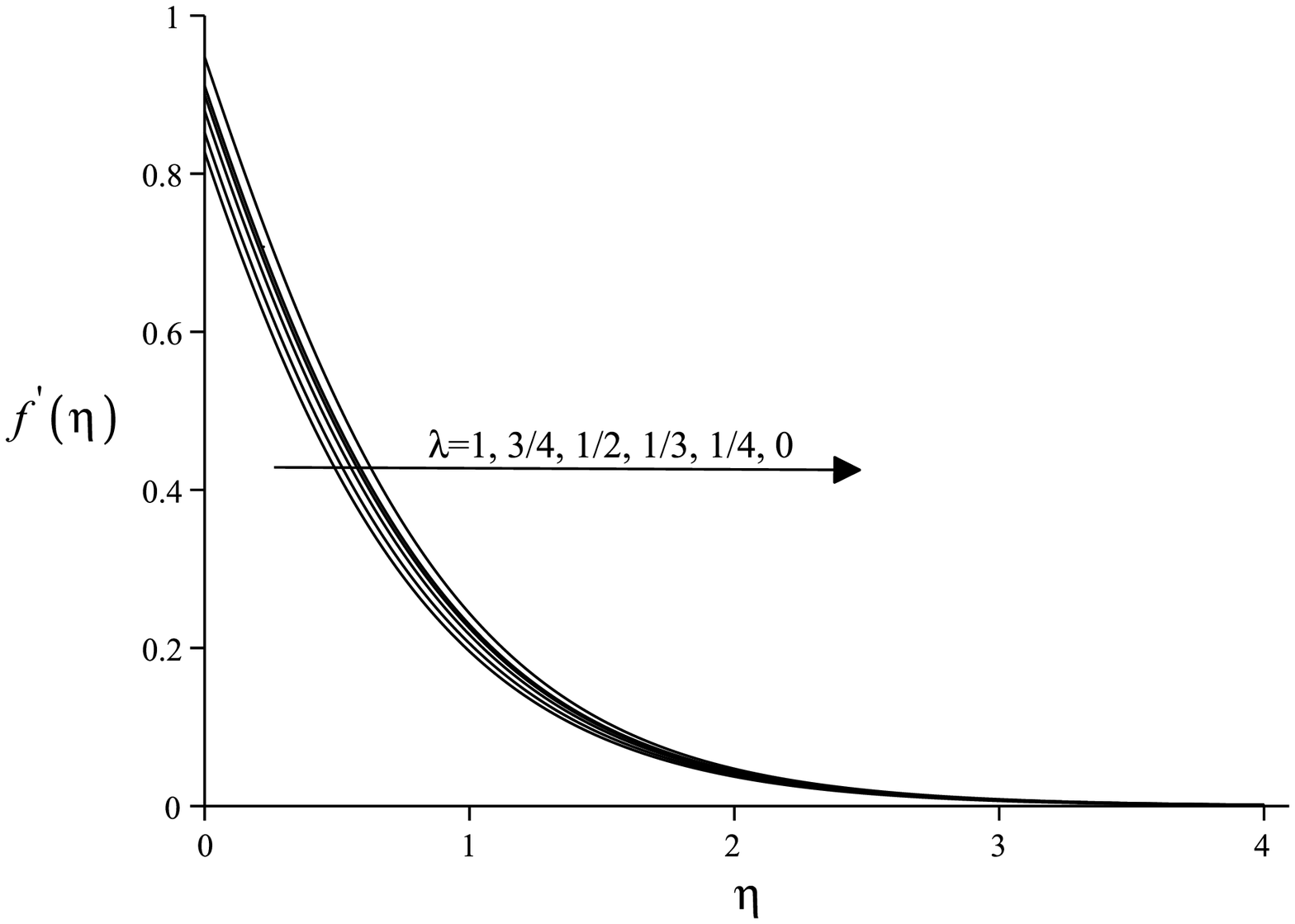}
\caption{IRBF approximation of $f'(\eta)$ for different values $\lambda=0,~1/4,~1/3,~1/2,~3/4$ and $1$}
\label{Fig.heat_flux_IRBF}
\end{figure}

\begin{figure}
\includegraphics[scale=0.5]{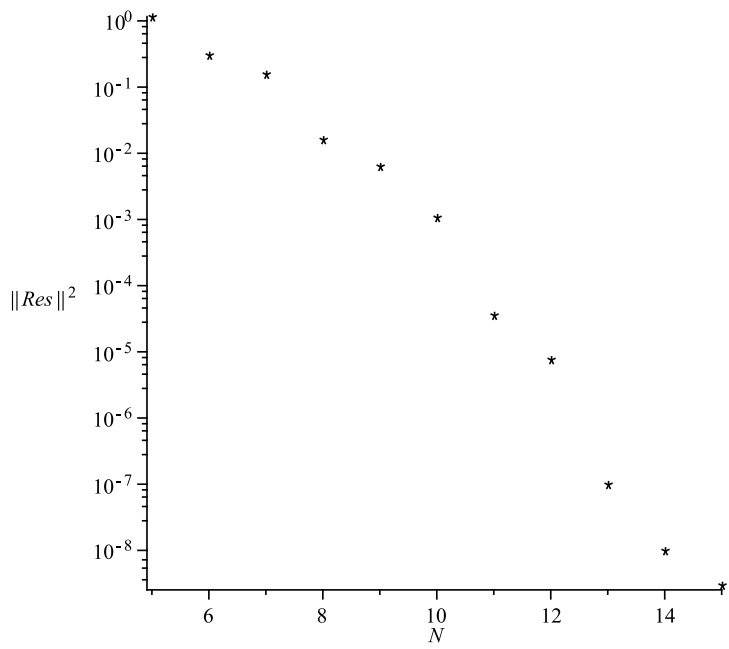}
\caption{$\|Res\|^2$ for $\lambda=2/3$}
\label{Fig.beta-2-3-Res}
\end{figure}

Table~(\ref{Tab.IRBF.Results.f'(0)}) shows the $f'(\eta)$ for some
$\lambda$ in comparison with solutions of \cite{Sohouli.Famouri}.
Also $f'(\eta)$ for two selected $\lambda=1/4$ and $3/4$ are
showed in Table~(\ref{Tab.IRBF.Results.f'(etta)}) in comparison with Runge-Kutta solution is obtained by the MATLAB software command ODE45 which is used and applied by the authors in ref. \cite{Sohouli.Famouri}. Absolute errors
show that IRBF give us approximate solution with a high degree of
accuracy with a small $N$. The resulting graph of
Eq.~(\ref{MainEquation}) is shown in Figure
(\ref{Fig.heat_flux_IRBF}). A graph in figures
(\ref{Fig.beta-2-3-Res}) for $\lambda=2/3$ show $\|Res\|^2$ for
some $N$. Table~(\ref{Tab.r.N}) shows $\|Res\|^2$ for some $N$ and
$\lambda$.\\
Comparison between DRBF solution in Table (\ref{Tab.DRBF.Results.f'(etta)}) with $N=12$ and IRBF solution in Table (\ref{Tab.IRBF.Results.f'(etta)}) with $N=10$ for $f'(\eta)$ show that the
convergence of the IRBF method is faster, because of using less numbers of collocation points.
\section{Conclusion}\label{Sec_Conclusion}
In this paper we made a comparison between the two common
collocation approaches based on radial basis functions namely DRBF
and IRBF methods on natural convection equation about an inverted
heated cone embedded in a porous medium of infinite extent which
are of great importance in the design of canisters for nuclear
wastes disposal. These functions are proposed to provide an
effective but simple way to improve the convergence of the
solution by collocation method. The direct approach (DRBF) is
based on a differentiation process, all derivatives are very
sensitive to noise arising from the interpolation of DRBFs from a
set of discrete data points. Any noise, even at the small level,
will be bad magnified with an increase in the order of derivative.
The indirect technique (IRBF) which is based on integration
process, each integral represents the area under the corresponding
curve, is much less sensitive to noise. Based on this observation,
it is expected that through the integration process, the
approximating functions are much smoother and therefore have
higher approximation power. Additionally, through the comparison
with other methods such as HAM we show that the RBFs methods have
good reliability and efficiency. Also high convergence rates and
good accuracy are obtained with the proposed method using
relatively low numbers of data points.

\clearpage
\listoffigures
\listoftables
\newpage
\clearpage

\clearpage

\clearpage
\begin{table}\small
\begin{tabular} {|l l|} \hline
\multicolumn{2}{|l|}{\textbf{Nomenclature}}\\
$A$                & prescribed constant\\
$f$                & similarity function for stream function temperature\\
$g$                & acceleration due to gravity parameter\\
$K$                & permeability of the fluid-saturated porous medium\\
$Nu_x$               & local Nusselt number \\
$q_w$              & surface heat flux fluid-saturated porous medium\\
$r$                & local radius of the cone fluid\\
$Ra_x$             & local Raleigh number\\
$T$                & temperature\\
$T_\infty$         & ambient temperature\\
$u,v$              & velocity vector along x,y axis\\
$x,y$              & Cartesian coordinate system\\
$x_0$              & distance of start point of cone from the vertex\\
\multicolumn{2}{|l|}{\textbf{Greek symbols}}\\
$\alpha$& thermal diffusivity the fluid-saturated porous medium\\
$\beta$   & expansion coefficient of the fluid\\
$\eta$    & independent dimensionless\\
$\theta$  & similarity function for\\
$\lambda$ & prescribed constants\\
$\mu$     & viscosity of the fluid\\
$\rho_\infty$ & density of the fluid at infinity\\
$\psi$    & stream function\\
\hline
\end{tabular} \label{table}
\end{table}
\clearpage
\begin{table}
\caption{Some well--known functions that generate RBFs $(r=\|x-x_{i}\|=r_{i}),~c>0$}
\begin{tabular}{l l}
\hline
Name of functions & Definition \\
\hline
Multiquadrics (MQ)  & $\sqrt{r^2+c^2}$\\
Inverse multiquadrics (IMQ) & $1/(\sqrt{r^2+c^2})$ \\
Thin plate (polyharmonic)Splines (TPS)  & $(-1)^{k+1}r^{2k}log(r)$ \\
Conical splines & $r^{2k+1}$ \\
Gaussian (GA) & $exp(-cr^2)$ \\
Exponential spline & $exp(-cr)$ \\
\hline
\end{tabular}
\label{Tab.RBR.definition}
\end{table}
\clearpage
\begin{table}
\caption{A comparison between solutions obtained by \cite{Sohouli.Famouri} and the DRBF method for $f'(0)$}
\begin{tabular*}{\columnwidth}{@{\extracolsep{\fill}}*{8}{c}}
\hline
\multicolumn{1}{c}{}&\multicolumn{1}{c}{Runge-Kutta}& \multicolumn{4}{c}{DRBF method}& \multicolumn{1}{c}{Other methods}\\
\cline{3-6} \cline{7-7}
$\lambda$ & Solution\cite{Sohouli.Famouri}  & N & c & DRBF & Error with RK &HAM\cite{Sohouli.Famouri} \\
\hline
$0  $ & $0.94760$ & $10$ & $3.46543$ & $0.94750$ & $0.0001$ & $0.94783$\\ 
$1/4$ & $0.91130$ & $12$ & $3.943$ & $0.91086 $ & $0.00044$ & $0.91119$\\ 
$1/3$ & $0.90030$ & $10$ & $4.9665$ & $0.90038$ & $0.0008$ & $0.90103$\\ 
$1/2$ & $0.87980$ & $10$ & $5.36$ & $0.87981$ & $0.00001$ & $0.87964$\\ 
$3/4$ & $0.85220$ & $12$ & $5.23$ & $0.85227$ & $0.00007$ & $0.85242$\\ 
$1  $ & $0.82760$ & $10$ & $5.89$ & $0.82737$ & $0.00023$ & $0.82726$\\ 
\hline
\end{tabular*}
\label{Tab.DRBF.Resultsf'(0)}
\end{table}
\begin{table}
\caption{Comparison between DRBF solution and Runge-Kutta solution for $f'(\eta)$ for $\lambda=1/4$ and $\lambda=3/4$ with $N=12$}
\begin{tabular*}{\columnwidth}{@{\extracolsep{\fill}}*{7}{c}}
\hline
\multicolumn{1}{c}{}& \multicolumn{3}{l}{$\lambda=1/4, f'(\eta)$}& \multicolumn{3}{l}{$\lambda=3/4, f'(\eta)$}\\
\cline{2-4} \cline{5-7}
$$    & DRBF      & Runge-Kutta & Error & DRBF      & Runge-Kutta  & Error \\
$\eta$& Solution & Solution\cite{Sohouli.Famouri}    & with RK   & Solution & Solution\cite{Sohouli.Famouri}     & with RK  \\
\hline
$  0$ & $0.910886$    & $0.911295$     & $0.000409$     &  $0.852268$    & $0.852193$     & $0.000075$\\
$0.1$ & $0.813122$    & $0.813604$     & $0.000482$     &  $0.755678$    & $0.755377$     & $0.000301$\\
$0.2$ & $0.720720$    & $0.721351$     & $0.000631$     &  $0.666229$    & $0.665448$     & $0.000781$\\
$0.3$ & $0.634656$    & $0.635531$     & $0.000875$     &  $0.584176$    & $0.582985$     & $0.001191$\\
$0.4$ & $0.555601$    & $0.556661$     & $0.001061$     &  $0.509628$    & $0.508141$     & $0.001487$\\
$0.5$ & $0.483877$    & $0.484997$     & $0.001120$     &  $0.442519$    & $0.440849$     & $0.001670$\\
$0.6$ & $0.419467$    & $0.420587$     & $0.001120$     &  $0.382620$    & $0.380907$     & $0.001713$\\
$0.7$ & $0.362196$    & $0.363276$     & $0.001080$     &  $0.329539$    & $0.327973$     & $0.001566$\\
$0.8$ & $0.311655$    & $0.312677$     & $0.001022$     &  $0.282878$    & $0.281536$     & $0.001342$\\
$0.9$ & $0.267358$    & $0.268264$     & $0.000906$     &  $0.242076$    & $0.241013$     & $0.001043$\\
$  1$ & $0.228756$    & $0.229508$     & $0.000752$     &  $0.206652$    & $0.205832$     & $0.000820$\\
$1.1$ & $0.195268$    & $0.195878$     & $0.000610$     &  $0.176015$    & $0.175434$     & $0.000581$\\
$1.2$ & $0.166342$    & $0.166847$     & $0.000505$     &  $0.149653$    & $0.149275$     & $0.000378$\\
$1.3$ & $0.141450$    & $0.141837$     & $0.000387$     &  $0.127026$    & $0.126821$     & $0.000205$\\
$1.4$ & $0.120090$    & $0.120362$     & $0.000272$     &  $0.107681$    & $0.107596$     & $0.000085$\\
$1.5$ & $0.101811$    & $0.102025$     & $0.000214$     &  $0.091167$    & $0.091196$     & $0.000029$\\
\hline
\end{tabular*}
\label{Tab.DRBF.Results.f'(etta)}
\end{table}
\clearpage
\begin{table}
\caption{A comparison between solutions obtained by \cite{Sohouli.Famouri} and the IRBF method for $f'(0)$}
\begin{tabular*}{\columnwidth}{@{\extracolsep{\fill}}*{8}{c}}
\hline
\multicolumn{1}{c}{}&\multicolumn{1}{c}{Runge-Kutta}& \multicolumn{4}{c}{IRBF method}& \multicolumn{1}{c}{Other methods}\\
\cline{3-6} \cline{7-7}
$\lambda$ & Solution\cite{Sohouli.Famouri}  & N & c & IRBF & Error with RK &HAM\cite{Sohouli.Famouri} \\
\hline
$0  $ & $0.94760$ & $10$ & $1.860$ & $0.94758$ & $0.00002$ & $0.94783$\\ 
$1/4$ & $0.91130$ & $10$ & $2.005$ & $0.91128$ & $0.00002$ & $0.91119$\\ 
$1/3$ & $0.90030$ & $10$ & $2.050$ & $0.90030$ & $0.00000$ & $0.90103$\\ 
$1/2$ & $0.87980$ & $10$ & $2.150$ & $0.87979$ & $0.00001$ & $0.87964$\\ 
$3/4$ & $0.85220$ & $10$ & $2.418$ & $0.85206$ & $0.00014$ & $0.85242$\\ 
$1  $ & $0.82760$ & $10$ & $2.380$ & $0.82762$ & $0.00002$ & $0.82726$\\ 
\hline
\end{tabular*}
\label{Tab.IRBF.Results.f'(0)}
\end{table}
\begin{table}
\caption{Comparison between IRBF solution and Runge-Kutta solution for $f'(\eta)$ for $\lambda=1/4$ and $\lambda=3/4$ with $N=10$}
\begin{tabular*}{\columnwidth}{@{\extracolsep{\fill}}*{7}{c}}
\hline
\multicolumn{1}{c}{}& \multicolumn{3}{l}{$\lambda=1/4, f'(\eta)$}& \multicolumn{3}{l}{$\lambda=3/4, f'(\eta)$}\\
\cline{2-4} \cline{5-7}
$$    & IRBF      & Runge-Kutta & Error & IRBF      & Runge-Kutta  & Error \\
$\eta$& Solution & Solution\cite{Sohouli.Famouri}    & with RK    & Solution & Solution\cite{Sohouli.Famouri}     & with RK    \\
\hline
$  0$ & $0.911278$    & $0.911295$     & $0.000017$     &  $0.852059$    & $0.852193$     & $0.000134$\\
$0.1$ & $0.813594$    & $0.813604$     & $0.000010$     &  $0.755254$    & $0.755377$     & $0.000123$\\
$0.2$ & $0.721302$    & $0.721351$     & $0.000049$     &  $0.665278$    & $0.665448$     & $0.000170$\\
$0.3$ & $0.635394$    & $0.635531$     & $0.000137$     &  $0.582736$    & $0.582985$     & $0.000249$\\
$0.4$ & $0.556440$    & $0.556661$     & $0.000221$     &  $0.507827$    & $0.508141$     & $0.000314$\\
$0.5$ & $0.484753$    & $0.484997$     & $0.000244$     &  $0.440510$    & $0.440849$     & $0.000339$\\
$0.6$ & $0.420334$    & $0.420587$     & $0.000253$     &  $0.380599$    & $0.380907$     & $0.000308$\\
$0.7$ & $0.362989$    & $0.363276$     & $0.000287$     &  $0.327647$    & $0.327973$     & $0.000326$\\
$0.8$ & $0.312347$    & $0.312677$     & $0.000330$     &  $0.281184$    & $0.281536$     & $0.000352$\\
$0.9$ & $0.267934$    & $0.268264$     & $0.000330$     &  $0.240696$    & $0.241013$     & $0.000317$\\
$  1$ & $0.229197$    & $0.229508$     & $0.000311$     &  $0.205482$    & $0.205832$     & $0.000350$\\
$1.1$ & $0.195584$    & $0.195878$     & $0.000294$     &  $0.175053$    & $0.175434$     & $0.000381$\\
$1.2$ & $0.166546$    & $0.166847$     & $0.000301$     &  $0.148891$    & $0.149275$     & $0.000384$\\
$1.3$ & $0.141548$    & $0.141837$     & $0.000289$     &  $0.126408$    & $0.126821$     & $0.000413$\\
$1.4$ & $0.120107$    & $0.120362$     & $0.000255$     &  $0.107158$    & $0.107596$     & $0.000438$\\
$1.5$ & $0.101771$    & $0.102025$     & $0.000254$     &  $0.090757$    & $0.091196$     & $0.000439$\\
$2  $ & $0.043769$    & $0.043951$     & $0.000182$     &  $0.038887$    & $0.039223$     & $0.000336$\\
$2.5$ & $0.018310$    & $0.018546$     & $0.000236$     &  $0.016225$    & $0.016574$     & $0.000349$\\
$3  $ & $0.007391$    & $0.007610$     & $0.000219$     &  $0.006433$    & $0.006832$     & $0.000399$\\
$3.5$ & $0.002716$    & $0.002953$     & $0.000237$     &  $0.002255$    & $0.002668$     & $0.000413$\\
$4  $ & $0.000719$    & $0.000962$     & $0.000243$     &  $0.000445$    & $0.000913$     & $0.000468$\\
$4.5$ & $0.000010$    & $0.000123$     & $0.000113$     &  $-0.00001$    & $0.000237$     & $0.000247$\\
\hline
\end{tabular*}
\label{Tab.IRBF.Results.f'(etta)}
\end{table}
\clearpage
\begin{table}
\caption{$\|Res\|^2$ for different $N$ and $\lambda$ by IRBF }
\begin{tabular*}{\columnwidth}{@{\extracolsep{\fill}}*{8}{c}}
\hline
$\lambda$ & N=5  & N=6 & N=8 & N=10 & N=12 &N=15 \\
\hline
$0  $ & $1.161037$ & $0.306257$ & $0.016229$ & $0.000108$ & $0.77e-5$ & $0.11e-7$\\ 
$1/4$ & $0.851976$ & $0.278068$ & $0.018364$ & $0.000363$ & $0.52e-4$ & $0.27e-6$\\ 
$1/3$ & $0.770207$ & $0.269714$ & $0.018842$ & $0.000494$ & $0.24e-4$ & $0.13e-5$\\ 
$1/2$ & $0.637068$ & $0.216095$ & $0.016813$ & $0.000587$ & $0.83e-5$ & $0.47e-6$\\ 
$3/4$ & $0.469112$ & $0.230425$ & $0.020280$ & $0.001051$ & $0.15e-4$ & $0.83e-6$\\ 
$1  $ & $0.355474$ & $0.221336$ & $0.020463$ & $0.001077$ & $0.31e-4$ & $0.39e-6$\\ 
\hline
\end{tabular*}
\label{Tab.r.N}
\end{table}
\end{document}